# Studying Paths of Participation in Viral Diffusion Process

Jarosław Jankowski[1], Sylwia Ciuberek[2], Anita Zbieg[3], Radosław Michalski[2]

[1] Faculty of Computer Science, West Pomeranian University of Technology, Szczecin, Poland
jjankowski@wi.zut.edu.pl
[2] Institute of Informatics, Wrocław University of Technology, Poland
{sylwia.ciuberek,radoslaw.michalski}@pwr.wroc.pl
[3] Institute of Psychology, University of Wrocław; Wroclaw University of Economics, Poland
anita.zbieg@gmail.com

**Abstract.** Authors propose a conceptual model of participation in viral diffusion process composed of four stages: awareness, infection, engagement and action. To verify the model it has been applied and studied in the virtual social chat environment settings. The study investigates the behavioural paths of actions that reflect the stages of participation in the diffusion and presents shortcuts, that lead to the final action – the attendance in a virtual event. The results show that the participation in each stage of the process increases the probability of reaching the final action. Nevertheless, the majority of users involved in the virtual event did not go through each stage of the process but followed the shortcuts. That suggests that the viral diffusion process is not necessarily a linear sequence of human actions but rather a dynamic system.

## 1 Introduction

The studies that direct attention to diffusion of innovation process [25], [7], [32], social influence mechanism [10], [21], [24], [12], [23] social contagion and epidemics outbreaks [5], [6] or cascades of influence patterns [34] investigate a similar phenomenon: a propagation, transmission and adoption of information (content, opinions, behaviours, emotions) within a network of social relations. As the information generated and shared online is gaining steadily in importance [28], more and more researchers are trying to deal with the power of electronic contagions. They are especially interested in social networking sites [26], [9] being recently the most popular online activity that has outnumbered e-mail actions [20], [1]. Because there is a great need to understand mechanisms and factors crucial for the spread of information, researchers search for new ways on how to study the phenomenon. The results from research areas related to dynamics and mechanisms of social transmission and adoption are successfully adopted to word of mouth process investigation [11], [33], [22] and viral seeding strategies examination [17], [4]. Nevertheless, the empirical network studies seem to be promising, relatively little has been done in this area [31].

The research presented in this paper is targeted to online social platforms with the ability to capture different forms of users' behaviours: communications, activity and transfers among users. Most of the research in the field of information diffusion and viral contagion is addressed to participants' characteristics, the structures of network they are embedded in or/and characteristics of information that is transmitted. Rare are studies in which attention is paid directly to the behaviours. The main motivation in the current research is to observe human action systems in more detailed way -by analysing different forms of behaviours related directly and indirectly to viral campaign and stages of the participation for the viral action. The study specifies several social behaviours related to the diffusion process and identifies several stages of participation in viral diffusion, starting from activity before receiving or sending viral information and behaviours after infection towards reaching a final diffusion goal

## 2 Related work

The process of information contagion among individuals and their further participation in particular viral actions can be observed and studied in three-folds[22]. Researchers focus on personal characteristics of people engaged in social contagion, their needs and motivations. Vital for the diffusion process are also social factors such as the characteristics of other people influencing an individual, attributes of the channels through which information flows and attributes of social system in which individuals operate.And finally, researchers consider the characteristics of spread information to be important for the social contagion.

### 2.1 Personal Characteristics and Stages of Adoption

Under the study there are personal characteristics of an individual who passes the message further as well as the one who is exposed to the message. What matters for the virus propagation are the personality traits like extraversion and innovativeness [4], authority of the sender, activity of the receiver [36] and similarity of sender and receiver demographics traits [29]. Moreover, people share information motivated by the need to be part of a group, but the need to be individualistic and stand out from a crowd is reported as a second reason [18]. Vital for virus propagation is also the need to be altruistic and the need for personal growth [18].

**Multistage Models of Engagement.**

The change of individual's opinion or behaviour is studied as the process of perceptional adaptation to a new stimuli, e.g. information, opinion, product or technology. The behavioural success of adoption depends on the cognitive processes that engage individual's attention as well as on personal motivations and emotions that lead to the interest. A few stages of the adoption have been distinguished [25], [8]. To adopt the received message, an individual first needs to pay attention or simply identify and notice information which is called the *awareness stage*. In the next stage of adoption (*interest stage*) any interest about the news is required. Interest leads to the engage-ment and active learning about the news by e.g. studying, searching, discussing, using or sharing it. Finally, the necessary information about the news, the decision of adoption or rejection is made (final *decision stage)*. The process was first reflected in the theory originally developed by Rogers [25] and successfully adopted and simplified to WOM marketing e-mail contagion research in the form presented above [8]. It has also been used as the background for the model and computer simulation of the decision to participate in viral marketing campaigns [33].

### 2.2    Social Factors and Mechanisms of Adoption

Nevertheless, the change of behaviour is preceded by intra-individual cognitive process, the perception and behaviour of a person is also influenced by inter-individual social context a person is embedded in. The innovation is communicated among the peers, friends, acquaintances and other members of a larger social group. As Rogers defines diffusion as "the communication of an innovation over time through certain channels among a social system" [25], the innovation is adopted due to interpersonal interaction within the social network.

**Social Influence, Imitation and Social Learning.**

Social learning theory [3] explains how people learn within a social context and assumes that the behaviour of an individual is influenced by other peoples' behaviour as well as by the personal characteristics of an individual. Moreover, behaviour of an individual influences the behaviour of others in a similar way that others influence an individual, what is called the reciprocal determinism. For the change of behaviour some important stages are required. First, an individual while observing a new behaviour needs to pay attention to characteristics of the behaviour to be able to imitate it. Next, the retention stage is needed, and this means that an individual is able to remember details of the behaviour. It is also possible for an individual to reproduce or imitate the behaviour and organize their own responses in accordance with the observed behaviour of others, which usually improves with practice. To do so, an individual must have the motivation or some other incentives [3]. Without any motivation, even if the previous stages are present, no imitation occurs. On the other hand, the social influence phenomenon occurs when people intentionally and directly influence others, being their friends, co-workers, family, acquaintances, etc. When other people act on the targeted individual, the closer they are to a person (physically or psychologically) and the more they are - the strongest social influence occurs [21], [24]. Individuals have several motivations to follow other people. Social diffusion operates through spreading awareness and interest for an information, social learning about the benefits and risks or normative influence extending the validity and legitimacy of the news. Informative influence occurs when there is lack of time or lack of other sources of knowledge and when the concerns that not adopting (when majority already adopted) can result in some disadvantages [30].

**The Number of Connections and the Number of Adopting Friends.**

A two-step flow model of communications [19] considered a small group of people called influentials as important for social influence and diffusion process, as they directly influence many neighbours. Influentials can be people with many connections occupying a central position in the social network [17], [15]. However, there are studies reporting that the viral content received from individuals with many ties have bigger chance to be ignored [22], [35], [13]. There is also a field of discussion if people with many connections are more or less susceptible in influencing other people. As they participate in a network more frequently, they are more often exposed to anything that flows through the network including a virus or a viral content [17], [6]. On the other hand, individuals with many friends were observed as less likely to be influenced by others and this is what researchers explain as generally weaker tie strength formed by highly connected individuals [2].

**Characteristics of Relations and Social Structures.**

Strength of the tie that connects two individuals is one of the most important characteristic having impact on the flow of information [22], [17], [4], [36]. Tie strength can be reflected by the number of common friends or triples formed by two individuals. Consumers are more likely to open e-mail messages sent from a person that they feel close to [8], [22] and stronger ties can increase the likelihood that the message will be passed along to others [26], [22]. Strong ties built on time spent together, emotional intensity, intimacy and reciprocal services between people [16], facilitate the flow of information. However, the weak ties serve as bridges creating short paths in the network [16] and allow us to reach a larger number of people in the network and to traverse greater social distance.

## 2.3 Characteristics of Transmitted Information

Vital for the awareness, usage and transmission of information is its relevance to the preferences of the receiver [22]. In Second Life community study [2], for the contagion it was vital, if the asset was popular or niche, as both types of assets were spreading through the network differently.

## 2.4 Limitation of earlier approaches and motivation

Majority of the presented studies concentrate on characteristics of people engaged in the diffusion process, channels and structures through which an information is transmitted and the characteristics of the information. Relatively small work is addressed to the behaviours of participants and social stimuli that directly or indirectly come from behaviour of other people engaged in the diffusion process. The goal of proposed model is to focus on those behaviours. Authors take into account micro (personal decision) and macro (social stimuli) perspective of human action in the context of viral diffusion process. The study captures participants' activity before infection, actions related to information spreading, activity after infection and reaching the goal of diffusion which in our case is the participation in the event.

While multistage models of participation in the diffusion of information attempt to fill the research gap and investigate participants behaviours, the observations of this kind focus mainly on the behavioural paths previously assumed theoretically and ignore other paths that can occur. The authors propose and verify the conceptual model of participation in the diffusion process, and the study is not limited to the observations of behavioural paths assumed in the model. The present work attempts to observe all possible behavioural paths that reflect the stages of engagement in the diffusion process and presents as well some behavioural shortcuts that lead to the main information diffusion goal which is the decision to participate in the event.

## 3      Conceptual framework

The model is targeted to virtual avatar chat world where the interactions among communicating users are more similar to a real world environment. Virtual versions of real goods can be used and exchanged. Interest can be built for example by observing goods possessed by other users or by messaging the information. In Fig. 1, the exemplary stages of communication based on viral content and typical activities in the system are shown.

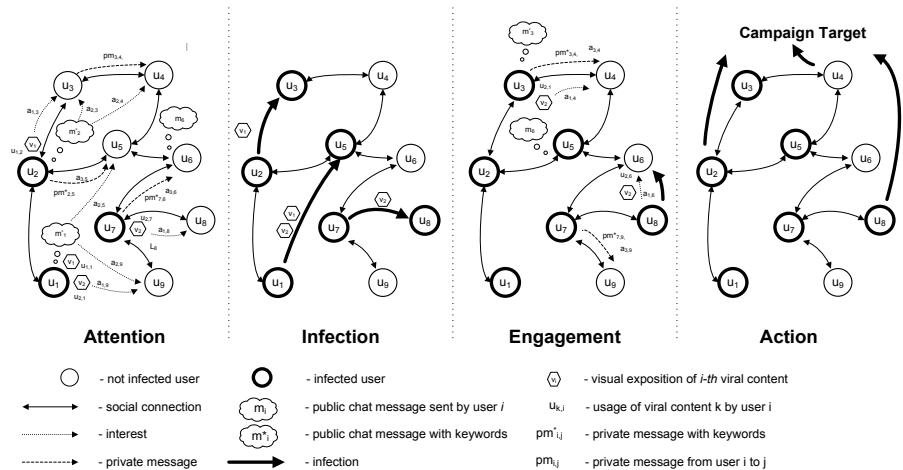

**Fig. 1.** Exemplary viral campaign in the online environment

Among users $U=\{u_1,u_2,u_3,u_4,u_5,u_6,u_7,u_8,u_9\}$ logged to the system, communication channels with messages $m$, $pm$ may be identified. At the *awareness stage,* users are exposed to different ways of building attention to viral campaign integrated with chat messages system. Both public and private messages can contain keywords related directly to viral action and are denoted as *m\** and *pm\** respectively. Communication containing keywords related directly to the campaign may build interest in action apart from the main viral visual components $v_1$ and $v_2$. Users $u_1, u_2, u_7$ are infected ear-lier and they have visual components in their repository. User $u_1$ has the set of two viral components $C_1=\{v_1,v_2\}$, assigned user $u_2$ has one component $C_2=\{v_1\}$ and user $u_7$ has one component $C_7=\{v_2\}$. Messages containing keywords and visual components related to the campaign may be exposed to other users, building their interest and demonstrating the engagement in action. At this stage there are factors building attention for the user: using visual objects $a_{1,i}$, public messages containing keywords $a_{2,i}$ and private messages with keywords $a_{3,i}$. At the *infection stage* users are infected with viral content: user $u_3$ receives content $v_1$ from user $u_2$, user $u_5$ receives content $v_1$ and $v_2$ from $u_1$, and user $u_8$ receives content $v_2$ from $u_7$. At the *engagement stage* infected users can engage in propagating information about campaign and increase their own interest in virus. Newly infected user $u_3$ exposes content $v_2$ to user $u_4$ increasing the current level of interest and sending private message $pm^*_{3,4}$ containing keyword used in the campaign. User $u_8$ exposes content $v_2$ to user $u_6$ and this results with the infection. User $u_5$ is not engaged in spreading the information about viral content and he/she sends public messages without any viral content. Finally, at Stage 4 (*action*) some of the infected users make up their decision and they reach the campaign target which can be defined as interaction required by the campaign organizer. Usually, only a fraction of users will reach this stage. In the exemplary process users $u_2$ and $u_8$ infected earlier with viral content, and user $u_4$ who was not infected with viral object but was exposed to the messages containing information about campaign, decided to perform the expected action. The example presented typical stages of communication from multiplayer online systems like games and virtual worlds. To generalize the approach, authors identified the characteristics of each phase of communication and characterized the process of participation in the campaign as composed of four stages: *attention*, *infection*, *engagement* and *action* presented in Fig. 2.

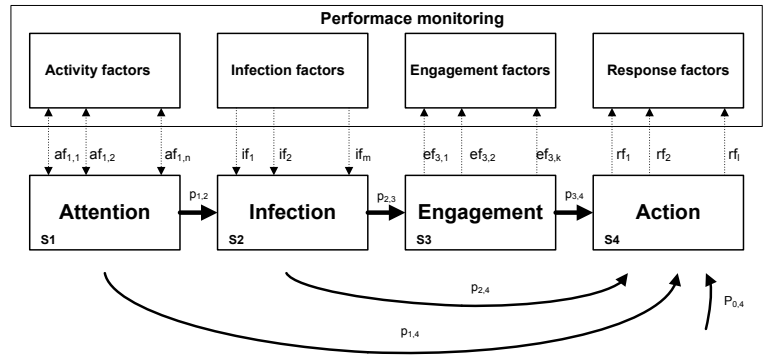

**Fig. 2.** A multistage model for viral campaign in online environment

At each stage there were factors specific to the particular stage. At the first stage related to building attention, a set of activity factors before infection $AF=\{af_{1,1},af_{1,2},\ldots,af_{1,n}\}$ is measured. At this stage, users can send and receive both public and private messages. Interest in action is measured as a ratio of messages containing keywords to all messages. Some of the users from the first stage can be infected with different components of viral action at the second stage. And at this stage infection factors can be measured from set $IF=\{if_{1,1}, if_{1,2},…,if_{1,n}\}$. After the infection took place, users can continue to engage themselves in the action but apart from messages with keywords they can spread viruses and use visual components to build the attention among other users. Factors measured at this level are included in a set $EF=\{ef_{1,1}, ef_{1,2},…,ef_{1,n}\}$. At the last stage response factors showing engagement in relation to the main goal of campaign may be identified. A set of response factors during campaign is denoted by $RF=\{rf_1, rf_2,…,rf_k\}$. Apart from users engaged in the campaign and viral transmission, the activity of users that were not infected by viral content is captured, as they "can go" to last stage without earlier infection. To gather information about users' activity, the transition probabilities between stages may be computed. Probability $p_{2,4}$ represents the probability of response without engagement after infection, $p_{3,4}$ represents the probability of reaching the goal of campaign after engaging in the spread of information about the campaign using different forms of activities. The $p_{0,4}$ denotes the probability of reaching the campaign goal without infection and no interest revealed in the campaign. Between stages the probabilities are measured as $p_{1,2}$ and $p_{2,3}$. Users do not necessarily need to take actively part in all stages of the campaign to perform the required action and in our case attend the event. In the system, the reference factor showing typical activity may be defined. Reference activity shows the typical action in the system as well as the overall activity in the period analyzed like number of logins, time spent in system or communication activity. The defined measures can be used for the evaluation of viral campaign not only in terms of viral spread but also as engagement in different stages of campaign, thus comparing different approaches. In the next step, the empirical research based on the proposed approach is presented.

## 4    Empirical results

To verify the proposed approach, authors used the dataset from online platform that connects functionality of multiplayer system and virtual world was used. The system combines function of chat and the entertainment platform where users are represented by graphical avatars. They have the opportunity to engage in the life of online community and to perform actions in different fields. Users have the access to virtual objects like avatars, clothes, different characters and special effects that can be distributed among users thought the viral transmission. The dataset was collected during the event with the main goal to motivate users to attend meeting in the system related to the virtual protest. The action was started five days prior to the event and two visual viral elements were used in a form of avatar face and informative object holding in avatar hand, which was an expression of a pro-protest attitude. The keywords related directly to viral action were identified. In the period $t$ there is the total number of 4910 unique users logged into the system. The viral content was initially delivered to 16 users selected from the most active group of protesters which became seeds for viral transmissions. The only way to obtain the items was by receiving it from other users.

To measure the campaign performance, a designed tracking system was used to capture all the actions related to the campaign by assigning them to the specific stages of campaign. The viral components were designed in a form of digital goods assigned to the user inventory. In the inventory, users can collect different elements of avatars or objects with special functionalities. After the viral infection, the digital object is added to the receiver inventory and it can be used in the visual chat environment. During the infection, the object is duplicated and a copy of the object is delivered to the receiver as well as stays at the sender's account. The tracking system was measuring the events related to sending objects and using the object at different stages of the process. The usage of the object was treated as an engagement and was represented by putting a mask on the avatar or using the transparent. Each time the mask or transparent was used, the system was saving the related actions with the assigned timestamps. Sequences of infections were used to build connection and to observe the diffusion of digital objects. Moreover, the tracking system was capturing keywords related to the campaign and assigning them as an additional measure of engagement. Also the information about users entering the dedicated chat room was collected. The obtained data was anonymized and no personal information about participants was used.

Among logged users 324 of them received viral content and 134 of them decided to send assets to friends which makes 41% of receivers engaged in forwarding messages. The monitoring of participation at each stage was based on the messages, usage of viral components and sending the viral content. The results showed that from all infected users 128 (39%) reached the goal of campaign and in the fifth day participated in the event. The event was visited by 197 users not infected directly by visual components. It shows that other factors affecting decisions and attitudes exist and not necessarily directly related to main viral campaign. The next part of the analysis was performed by using participation factors for the proposed multistage approach. The analysis was performed for activity within five days frame. To monitor engagement at this stage, factors showing communication activity containing keywords in relation to all messages sent by user during the monitoring period were defined. As observed in the campaign, most of the users' activity was recorded in the system and it was possible to track the activity before the infections. In the stage S1 (*awareness stage*) activity prior to infections is included, S2 (*infection stage*) contains activity related to infections and stage S3 (*engagement stage*) contains activity after infection prior to main event. Within stage S4 (*actions stage*) factors representing visiting event and activity during event where identified. The probabilities of transitions between stages where computed. Table 1 shows identified activity factors for all stages.

**Table 1.** Activity factors measured during campaign

| Factor | Description |
|---|---|
| S1:$af_{1,1}$ | Messages containing keywords sent by non-infected users before event divided by all sent messages |
| S1: $af_{1,2}$ | Messages with keywords received before event by non-infected users divided by number of messages received in this period |
| S1: $af_{1,3}$ | Messages containing keywords received by infected users before infection divided by all messages prior to infection |
| S1: $af_{1,4}$ | Messages containing keywords sent before infection by infected users divided by all messages prior to infection |
| S2: $if_{2,1}$ | Viral component received |
| S2: $if_{2,2}$ | Viral component sent |
| S3:$ef_{3,1}$ | Messages sent with the keyword after infection before event divided by all messages sent in this period |
| S3:$ef_{3,2}$ | Messages with keywords received after infection before event divided by number of messages received in this period |
| S3:$ef_{3,3}$ | Viral component used after infection before event divided by all messages sent by user |
| S4:$rf_1$ | Visit to event |
| S4:$rf_2$ | Number of messages sent during event by infected users with keywords divided by all messages sent during event |
| S4: $rf_3$ | Number of messages sent during event by non-infected users with keywords divided by all messages sent during event |
| S4:$rf_4$ | Number of times viral component was used during event divided by number of messages sent by user during event |

Among all logged users 1298 communicated the keywords and 245 of them were infected ($p_{1,2}$= 0.19). 152 users after infection sent messages containing keywords($p_{2,3}$= 0.47). 73 of the users from Stage 3 visited the event, resulting in $p_{3,4}$=0.48. A number of 1021 users among not infected used keywords in messages and 118 of them visited the event ($p_{1,4}$=0.12). 106 users after infections did not use keywords in messages and 44 of them visited the event, resulting in $p_{2,4}$=0.41. 125 users after infection sent messages with keywords and 61 of them visited the event ($p_{3,4}$=0.49). 3604 users (logged to system during five days) did not use keywords in messages and 79 of them visited event ($p_0$=0.022). Fig. 3 shows the probabilities assigned to each stage of the process.

The transitions between the stages show that reaching each level of the process increases the probability of attending the main event. The probability of attending the event was increased from $p_0$=0.022 to $p_{1,4}$=0.12 if the engagement in sending messages was observed even without infection. Receiving the viral content increased the probability of reaching the final stage to $p_{2,4}$=0.41 even if no engagement was observed after the infection. It shows that the viral content was affecting the decision to join event and the probability increased after engagement and infection to $p_{3,4}$=0.48.

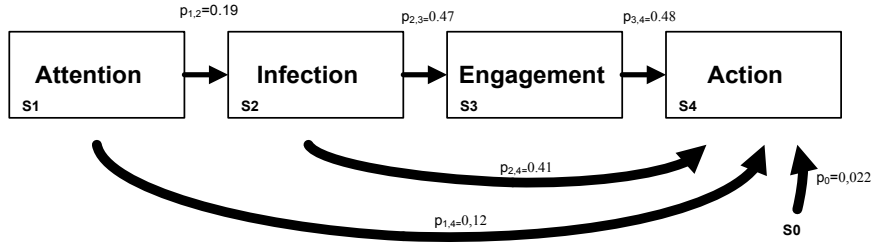

**Fig. 3.** Probabilities assigned to moving between stages.

The differences between users of group G1 visiting the event and group G2 not visiting the event among all users logged to the system within the monitored period were found. A number of 4585 users did not visit the event while 325 visited. Table 2 shows the analysis based on the Mann-Whitney significance test.

**Table 2.** Differences between visitors and non-visitors for group G1 and G2

| Factor | Rank Sum | Rank Sum | U | Z | p-value | Valid N | Valid N |
|---|---|---|---|---|---|---|---|
| $if_{2,1}$ | 1069235 | 10987270 | 473865.0 | 10.98194 | 0.000000 | 325 | 4585 |
| $ef_{3,2}$ | 1011660 | 11044846 | 531440.5 | 8.65046 | 0.000000 | 325 | 4585 |
| $af_{1,4}$ | 984894 | 11071612 | 558206.5 | 7.56659 | 0.000000 | 325 | 4585 |
| $ef_{3,3}$ | 982105 | 11074400 | 560995.0 | 7.45367 | 0.000000 | 325 | 4585 |
| $ef_{3,1}$ | 964057 | 11092449 | 579043.5 | 6.72281 | 0.000000 | 325 | 4585 |
| $if_{2,2}$ | 940898 | 11115608 | 602202.5 | 5.78500 | 0.000000 | 325 | 4585 |
| $af_{1,3}$ | 902094 | 11154412 | 641006.5 | 4.21366 | 0.000025 | 325 | 4585 |
| $af_{1,1}$ | 880253 | 11176252 | 662847.0 | 3.32924 | 0.000871 | 325 | 4585 |
| $af_{1,2}$ | 840438 | 11216067 | 702662.0 | 1.71696 | 0.085987 | 325 | 4585 |

Significant factors were related to the messages: received messages with keywords related to event after infection $ef_{3,2}$ and sending messages with keywords $af_{1,4}$ before infection. Before infection users interested in action were discussing about it and keywords related to event were used at this stage. Analysis show that engagement at early stage was resulting in higher interest to the main action. The next important factor $ef_{3,3}$ is showing that users in group G1 were more actively using viral components after the infection. The engagement in the event and factors affecting intensity of messages containing keywords during the event were also analysed. The results based on multiple regression are showed in Table 3.

Table 3. Factors affecting engagement during event

| Factor | b* | Std. Err. | b | Std. Err. | t(315) | p-value |
|---|---|---|---|---|---|---|
| Intercept | | | 0.067661 | 0.014063 | 4.811124 | 0.000012 |
| $ef_{3,3}$ | 0.000578 | 0.062929 | 0.000953 | 0.103667 | 0.009191 | 0.992673 |
| $ef_{3,2}$ | -0.001277 | 0.063362 | -0.001440 | 0.071463 | -0.020152 | 0.983935 |
| $ef_{3,1}$ | 0.070580 | 0.054658 | 0.019943 | 0.015444 | 1.291298 | 0.197547 |
| $if_{2,2}$ | 0.012411 | 0.062075 | 0.000573 | 0.002867 | 0.199935 | 0.841661 |
| $if_{2,1}$ | 0.092130 | 0.067792 | 0.010677 | 0.007856 | 1.359003 | 0.175118 |
| $af_{1,4}$ | **0.295254** | **0.059231** | **0.615257** | **0.123426** | **4.984810** | **0.000023** |
| $af_{1,3}$ | -0.052056 | 0.053760 | -0.123527 | 0.127568 | -0.968319 | 0.333628 |
| $af_{1,2}$ | -0.045759 | 0.050451 | -0.508764 | 0.560928 | -0.907004 | 0.365098 |
| $af_{1,1}$ | **0.316411** | **0.050446** | **0.906451** | **0.144517** | **6.272283** | **0.000012** |

Two from measured factors resulted as statistically significant. As the most important factor $af_{1,1}$ was identified, representing the number of messages with the keywords sent by non-infected users in relation to all messages before the event. The second important factor was $af_{1,4}$ showing the engagement in the distribution of messages with the keywords by infected users before the infection. In the next stage, multiple regression analysis was performed only using infected users and their activity during the event. As factors affecting the engagement the results showed $af_{1,4}$ and $ef_{3,1}$, representing the engagement in the action before the event. The engagement before the infection was more important than after the infection. And finally, no statistically significant factors were identified as affecting $af_4$ and the usage of visual components during the event. This result confirms that users engaged in the first stages of the process were more willing to engage in the final event.

## 5     Discussion

The performed experimental studies showed that reaching each level of the diffusion process increased the probability of attending the main event. The behaviour before the event (receiving and spreading messages and viral components) was related directly to the activity during the event. Not only the number of infections received by user was increasing its interest in action but also the number of infections sent by a user. Using viral components by a person increased the interest in action of people being receivers of the information, as well as confirms sender's interest in action as the expression of some attitude (e.g. manifesting the opinion or convincing others to it) strengthens the attitude. Interestingly, users directly interested in the action even without receiving the viral components joined it. Probably because the diffused information was not specific for the studied online platform and diffused also in a real world. Hence, the behaviour of some participants were not necessarily influenced by stimuli that directly came from other users but have their own attitude acquired outside the platform.

Tracking the steps of participation in the diffusion process, authors tried to fit users' behaviour to the AIEA model, which describes the increase of participation leading to the final stage – action. Generally, the basic intuition suggests that the users which were the most engaged would follow to the action stage. As the results showed on Figure 3, this kind of intuition might be misleading because this group represents only a small fraction of all users in the last stage. This analysis showed that there is a significant importance of other paths in the model which may be understood as shortcuts from one of the previous stages towards the last stage. For instance, a high fraction of users took part in the event who were registered previously only in the attention stage (S0 -> S4).In that case all the paths should be analysed with equal importance because the analysis showed that the majority of users participating in the final event were not necessarily the most engaged ones. This leads to the conclusion that in the online environment, the action performed by users may result in bigger changes if someone was not present in all stages. The other question arises: what is the role of users who are very engaged but not participating in the event? Although someone may ignore them, their role is invaluable in terms of influencing other users – if they would be passive in terms of infecting other users, the number of users in previous stages would be significantly lower. The vital role of users in the third stage is to increase the overall interest of users into the event and, as the analysis showed, even weak engagement of users in the whole multistage process will convert to the final action.

## 6      Summary

Social contagion studies are usually a simplification of a real world situation where decisions related to participation in viral diffusion are based on many different factors which are difficult to observe and monitor. The current study refers to some of them: verbal communication, visual aspects of infections when viral content is visible, interest to viral content revealed prior to infection and talks about it. The proposed model attempts to merge micro and macro dynamics of human behaviour. The former reflect intra-personal process of making the successive actions by a person, the latter pay attention to the inter-actions that occur between a person and other people.

The study showed that the stages in viral diffusion process are not necessarily sequential. By studying the additional paths (shortcuts) in the users' participation processes, the analysis showed that users mostly did not followed the basic path. That leads to the finding that viral diffusion process probably is not a linear sequence of actions but rather a dynamic system, while many models of diffusion assumes the infection as a part of a sequential process and many empirical studies covers only the observation of previously defined sequence of behaviours. That of course doesn't mean that we cannot find any sequential model that describes great a social phenomenon, but that empirical studies of social diffusion conducted with the interest to the dynamics of human actions can be an interesting way of further research. The pro-posed model and analysis may be used in empirical studies related to social diffusion process in online environments, as the studied functionality is similar to many popular online setting and the studied phenomenon is not only relevant to social informatics but can be applied to many fields related to social contagion: word of mouth and viral marketing, the studies of public opinion and actions, information systems and organizational research.

**Acknowledgments.** This work was partially supported by fellowship co-financed by the European Union within the European Social Fund, the Polish Ministry of Science and Higher Education, the research project 2010-13.

# References

1. Arndt, J.: Role of Product-Related conversations in the diffusion of a new product. Journal of Marketing Research, vol. 4, no. 3, pp. 291-295 (1967)
2. Bakshy, E., Karrer, B., Adamic, L.A.: Social influence and the diffusion of user-created content. In: Proceedings of the 10th ACM conference on Electronic commerce, ser. EC '09. New York, NY, USA: ACM, pp. 325-334 (2009)
3. Bandura, A.. Social cognitive theory of mass communication. In: J. Bryant, D. Zillman (eds.), Media effects: Advances in theory and research, Mahwah, NJ: Lawrence Erlbaum Associates. pp. 121–153 (2002)
4. Chiu, H.C., Hsieh, Y.C., Kao, Y.H., Lee, M.: The Determinants of Email Receivers' Disseminating Behaviors on the Internet. Journal of Advertising Research, vol.47, no. 3, pp. 524-534 (2007)
5. Christakis¸ N.A., Fowler, J.H.: The spread of obesity in a large social network over 32 years. N Engl J Med, vol. 357, no. 4, pp. 370-379 (2007)
6. Christakis N.A., Fowler, J.H.: Social network sensors for early detection of contagious outbreaks. PLoS ONE, vol. 5, no. 9, pp. e12 948+ (2010)
7. Coleman, J., Katz, E., Menzel, H.: The diffusion of an innovation among physicians. Sociometry, vol. 20, no. 4, pp. 253-270 (1957)
8. De Bruyn, A., Lilien, G.L.: A multi-stage model of word-of-mouth influence through viral marketing. International Journal of Research in Marketing, vol. 25, no. 3, pp. 151-163 (2008)
9. Emarketer Press Releases: Social Network Ad Revenues to Reach $10 Billion Worldwide in 2013 (2011)
10. Festinger, L.: Informal Social Communication. Psychological Review, vol. 57, no.5, pp. 271-282 (1950)
11. Godes,D., Mayzlin, D. : Firm-Created Word-of-Mouth communication: Evidence from a field test, Marketing Science, vol. 28, no. 4, pp. mksc.1080.0444-739(2009)
12. Goel, S., Mason, W., Watts, D.J.: Real and Perceived Attitude Agreement in Social Networks. Journal of Personality and Social Psychology, vol . 99, no. 4, pp. 611-621 (2010)
13. Goel, S., Watts, D.J., Goldstein, D.G.: The Structure of Online Diffusion Networks. forthcoming in ACM EC'12, (2012)
14. Golan, G.J.,Zaidner, L. : Creative Strategies in Viral Advertising: An Application of Taylor's Six-Segment Message Strategy Wheel. Journal of Computer-Mediated Communication, vol. 13, no. 4, pp. 959–972 (2008)

15. Goldenberg, J., Han, S., Lehmann, D.R., Hong, J.W.: The Role of Hubs in the Adoption Process. Journal of Marketing, vol. 73, no. 2, pp. 1-13 (2009)
16. Granovetter, M.: The Strength of Weak Ties: A Network Theory Revisited. Sociological Theory 1, pp. 201–233 (1983)
17. Hinz O., Skiera, B., Barrot, Ch., Becker, J.U.: Seeding Strategies for Viral Marketing: An Empirical Comparison. Forthcoming in Journal of Marketing, (2012)
18. Ho, J.Y.C., Dempsey, M. : Viral Marketing: Motivations to Forward Online Content. Journal of Business Research, vol. 63, no. 9/10, pp. 1000-1006 (2010)
19. Katz, E., Lazarsfeld, P: Personal Influence: The Part Played by People in the Flow of Mass Communications. Transaction Publishers (2005)
20. Keenan, A., Shiri, A.: Sociability and social interaction on social networking websites, Library Review, vol. 58, no. 6, pp. 438 - 450 (2009)
21. Latané, B.: The psychology of social impact. American Psychologist, vol. 36, no. 4, pp. 343-356 (1981)
22. Liu-Thompkins, Y.: Seeding viral content: Lessons from the diffusion of online videos. Forthcoming in Journal of Advertising Research, (2011)
23. McPherson, M., Smith-Lovin, L., Cook, J.M.: Birds of a Feather: Homophily in Social Networks. Annual Review of Sociology 27, pp. 415-444 (2001)
24. Nowak, A., Szamrej, J., Latané, B., Nowak, A., Szamrej, J., Latan, B.: From private attitude to public opinion: A dynamic theory of social impact. Psychological Review 97, pp. 362-376 (1990)
25. Rogers, E.M., Diffusion of Innovations, 5th ed. New York: Free Press, (2003)
26. Shu-Chuan, C., Kim, Y.: Determinants of consumer engagement in electronic word-of-mouth (eWOM) in social networking sites. International Journal of Advertising, vol. 30, no. 1, pp. 47-75 (2011)
27. Tarde G.: On Communication and Social Influence: Selected Papers. Heritage of Sociology Series, University Of Chicago Press, Reprint edition Apr. 15, 2011, first edition 1893.
28. Thackeray, R., Neiger, B. L., Hanson, C. L., McKenzie, J. F.: Enhancing Promotional Strategies Within Social Marketing Programs: Use of Web 2.0 Social Media. Health Promotion Practice, vol. 9, no. 4, pp. 338-343(2008)
29. Trusov, M., Bodapati, A.V., Bucklin,R.E.: Determining Influential Users in Internet Social Networks. Journal of Marketing Research, vol. 47, no. 4, pp. 643-658 (2010)
30. van den Bulte, Ch., Lilien, G.L.: Medical Innovation Revisited: Social Contagion versus Marketing Effort. American Journal of Sociology, vol. 106, no. 5, pp. 1409–35 (2001)
31. van den Bulte, Ch., Wuyts, S., Social Networks and Marketing. Cambridge, MA: Marketing Science Institute (2007)
32. Valente, T.W.: Network Models of the Diffusion of Innovations, Hampton Press, NJ, (1995)
33. van der Lans, R., G. van Bruggen, G., Eliashberg, J., Wierenga, B.: A Viral Branching Model for Predicting the Spread of Electronic Word of Mouth, Marketing Science, vol. 29, no. 2, pp. 348-365 (2010)
34. Watts, D.J.: A simple model of global cascades on random networks. Proceedings of the National Academy of Sciences, vol. 99, no. 9, pp. 5766-5771 (2002)
35. Watts, D.J., Dodds, P.S.: Influentials, networks, and public opinion formation. Journal of Consumer Research, vol. 34, no. 4, pp. 441-458 (2007)
36. Zbieg, A., Żak, B., Jankowski, J., Michalski, R., Ciuberek, S.: Studying Diffusion of Viral Content at Dyadic Level. Proceedings of the 2012 International Conference on Advances in Social Networks Analysis and Mining (ASONAM), IEEE Conference Publications, pp. 1291-1297 (2012)